\pdfoutput=1
\documentclass[aps,prl,twocolumn, titlepage,showpacs]{revtex4}

\usepackage{graphicx}
\usepackage{color}
\usepackage{dcolumn}

\usepackage{bm}

\begin{document}

\title{Observation of temperature peaks due to strong viscous heating in a dusty plasma flow}

\author{Yan Feng}
\email{yanfengui@gmail.com}
\author{J. Goree}
\author{Bin Liu}
\affiliation{Department of Physics and Astronomy, The University
of Iowa, Iowa City, Iowa 52242, USA}

\date{\today}

\begin{abstract}

Profound temperature peaks are observed in regions of high velocity shear in a 2D dusty plasma experiment with laser-driven flow. These are attributed to viscous heating, which occurs due to collisional scattering in a shear flow. Using measurements of viscosity, thermal conductivity, and spatial profiles of flow velocity and temperature, we determine three dimensionless numbers: Brinkman Br = 0.5, Prandtl Pr = 0.09, and Eckert Ec = 5.7. The large value of Br indicates significant viscous heating that is consistent with the observed temperature peaks.

\end{abstract}

\narrowtext

\maketitle

Viscous heating occurs when there is a shear in the velocity profile in a fluid. Viscous heating is the energy lost whenever there is a gradient in flow velocity, for example due to moving solid surfaces such as pistons in an oil-lubricated engine, a human swimming in water, or an airplane flying through air. Despite the importance of this energy loss, viscous heating often does not result in a measurable increase in temperature, because most fluids conduct heat rapidly away from the location of flow gradients where the heat is generated. This competition between viscous heating and thermal conduction is measured by the dimensionless parameter of fluid dynamics called the Brinkman number~\cite{Huba:94, Mills:95},
\begin{equation}\label{Brinkman}
{{\rm Br} = \frac{\eta}{\kappa} \frac{(\Delta v)^2}{\Delta T}.}
\end{equation}
Here, $\eta$ is the shear viscosity and $\kappa$ is the thermal conductivity~\cite{transport}, while $\Delta v$ and $\Delta T$ are differences in flow velocity and temperature between two positions, for example between a boundary and the hottest point within the flow. Note that ${\rm Br}$ depends on two instrinsic properties of the substance, viscosity $\eta$ and thermal conductivity $\kappa$, as well as properties of the flow profile $\Delta v$ and $\Delta T$. A substantial temperature increase due to viscous heating would be indicated by a value of Br of order unity~\cite{Mills:95}. While a value as large as unity is theoretically achievable in some simple flows, such a large value is seldom measured experimentally. Most commonly, Br is orders of magnitude smaller than unity~\cite{Tso:00, White:00}.

The Brinkman number can also be expressed as the product ${\rm Br} = {\rm Pr}\,{\rm Ec}$ of two other dimensionless numbers: the Prandtl number~\cite{White:00} ${\rm Pr}$ and Eckert number~\cite{Gschwendtner:04} ${\rm Ec}$, which are defined as
\begin{equation}\label{Prandtl}
{{\rm Pr} = \frac{\nu}{\chi} \equiv \frac{\eta/\rho}{\kappa/(c\rho)}}
\end{equation}
and
\begin{equation}\label{Eckert}
{{\rm Ec} = \frac{(\Delta v)^2}{c\, \Delta T}.}
\end{equation}
Here, $\nu \equiv \eta / \rho$ is the momentum diffusivity (i.e., kinematic viscosity), $\chi \equiv \kappa / (c \rho)$ is the thermal diffusivity, $c$ is the specific heat, and $\rho$ is the mass density. The Prandtl number is essentially a ratio of momentum diffusivity to energy diffusivity~\cite{White:00}, while the Eckert number characterizes dissipation as the ratio of a directed kinetic energy to a thermal kinetic energy~\cite{Gschwendtner:04}.

The meaning of the Brinkman number is motivated by the energy continuity equation~\cite{Batchelor:67, Landau:87} for steady conditions,
\begin{equation}\label{energy}
{\frac{\eta}{\rho} \left(\frac{\partial {v_x}}{\partial y}\right)^2 + \frac {\kappa}{\rho} \nabla^2 T + P_{\rm ext} = 0,}
\end{equation}
written here for the simplified geometry of a Couette flow. A Couette flow~\cite{Liepmann:57} is a laminar flow between two planar boundaries moving in the $x$ direction at different speeds $\Delta v$. The Brinkman number reflects the ratio of viscous heating to thermal conduction, i.e., the ratio of the first two terms in Eq.~(\ref{energy}). The last term $P_{\rm ext}$ represents any external source or sink of heat energy per unit mass.

Although the Brinkman number is usually tiny, we will show that it has a large value for the substance used in our experiment: a dusty plasma. A dusty plasma is a four-component mixture of micron-size particles of solid matter, electrons, positive ions, and a rarefied gas of neutral atoms~\cite{Melzer:08, Morfill:09, Piel:10, Shukla:02, Bonitz:10}. The small particles, which we term dust particles, gain a high negative electric charge $Q$, so that they collide with one another electrically, like charged colloids in an aqueous suspension~\cite{Lindsay:82}. The charged dust particles can self-organize into a lattice with crystalline properties, which can be melted to make a liquid~\cite{Chu:94, Thomas:96, Melzer:96, Sheridan:08, Samsonov:04, Hartmann:10, Feng:11}. A dusty plasma can be manipulated by external forces, which can account for $P_{\rm ext}$ in Eq.~(\ref{energy}), to cause a laminar flow of dust particles~\cite{Nosenko:04}.

Compared with other substances, dusty plasma has several extreme properties. The typical interparticle spacing is much larger than the particle size, about $0.5~{\rm mm}$ as compared to $8.09~{\rm \mu m}$ for our experiment, and correspondingly the volume fraction of solid material is extremely small, of order $10^{-6}$. Due to this low volume fraction, dusty plasma is one of the softest substances known; judged by its shear modulus when in crystalline form~\cite{Lindsay:82}, it is nineteen orders of magnitude softer than metals, and six orders of magnitude softer than colloidal crystals~\cite{Melzer:08}. Because of its extreme softness, a crystalline dusty plasma has a sound speed of only a few mm/s~\cite{Feng:10}, and it is possible to generate velocity gradients as large as the thermal velocity divided by the interparticle spacing, as we will show here. While a dusty plasma has a very small viscosity $\eta$~\cite{Nosenko:04}, it also has a very low mass density $\rho$, so that the kinematic viscosity $\nu \equiv \eta / \rho$ is actually comparable to that of water, $1~{\rm mm^2/s}$~\cite{Nosenko:04}.

With these extreme properties, it is worth asking whether significant viscous heating and a large value of ${\rm Br}$ can be attained in dusty plasma, and in this Letter we find that this is so. Our experiment is designed to yield all the parameters required to determine ${\rm Br}$, ${\rm Pr}$, and ${\rm Ec}$. Our measurements include a simultaneous quantification of the velocity profile $v_x(y)$, a profile of a kinetic temperature, and values for two transport coefficients, $\eta$ and $\kappa$.

\begin{figure}[htb]
\centering
\includegraphics{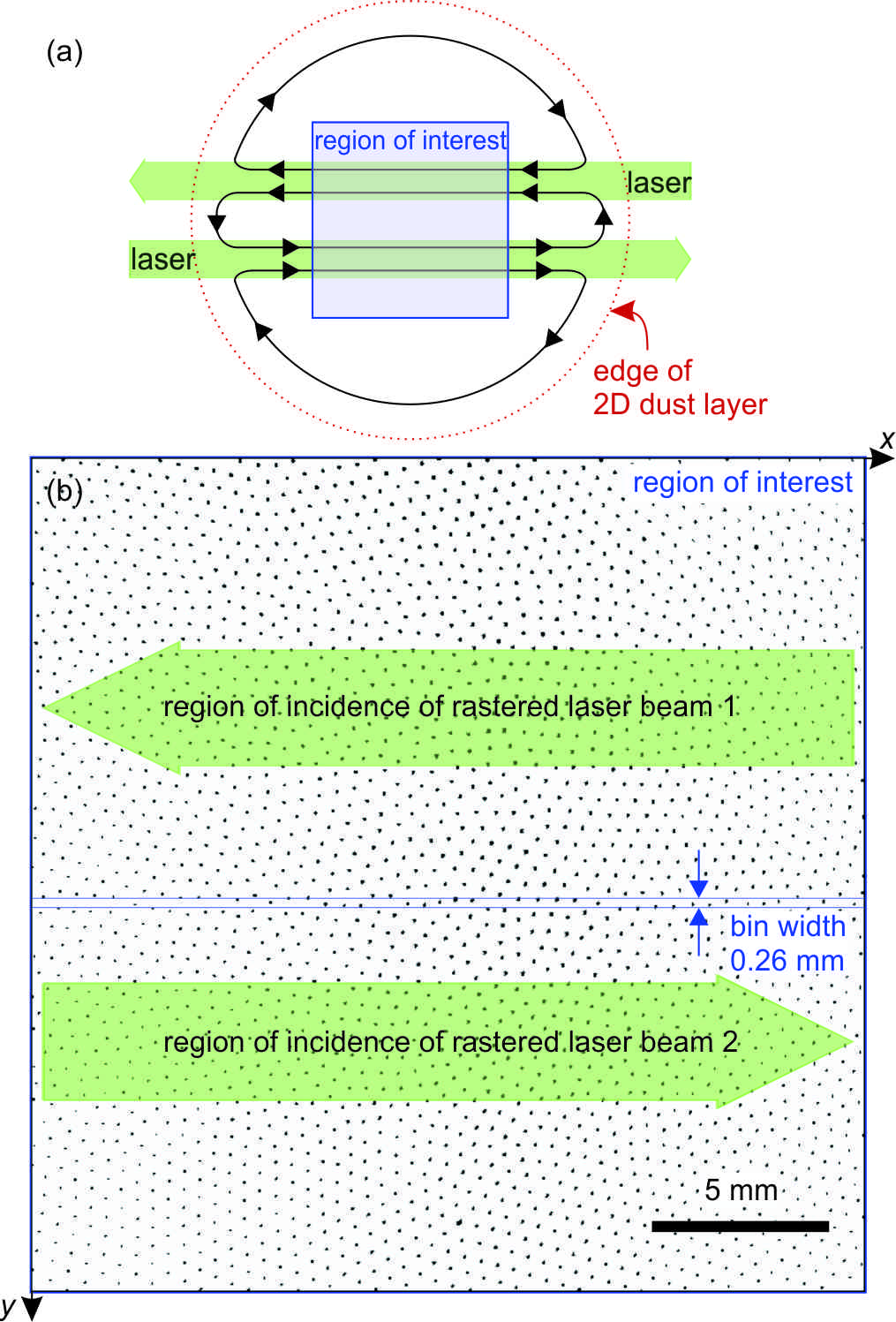}
\caption{\label{etasketch} (Color online). Sketch of laser-driven flows in the 2D dusty plasma. In the region of interest, the flow is straight, with curved flow limited to the edge of the dust layer, as sketched in (a). A video image of the dust particles within the region of interest is shown in (b). In our data analysis, we exploit the symmetry of the experiment, and divide the region of interest into 89 long narrow rectangular bins, with one example bin shown in (b). The particle data are converted to continuum data by averaging data for individual particles in each bin.}
\end{figure}

Our experimental setup was similar to the one described in~\cite{Feng:11} for another experiment. A vacuum chamber was filled with $15.5~{\rm mTorr}$ argon gas at room temperature, which was partially ionized by applying $214~{\rm V}$ peak-to-peak voltage at $13.56~{\rm MHz}$. About $10^4$ dust particles, which were $8.09~{\rm \mu m}$ diameter melamine-formaldehyde microspheres with a mass of $m_d = 4.18 \times 10^{-13}\,{\rm kg}$, were introduced into the plasma. They gained a large negative charge of $-9700\,e$ where $e$ is the elementary charge as determined by a wave-spectra method~\cite{Nunomura:02} which has an accuracy of about $\pm 10\%$. They became electrically levitated by a vertical dc electric field above a horizontal electrode due to its dc self-bias of $- 138~{\rm V}$. A weaker radial dc electric field in the plasma prevented the charged dust particles from escaping in the horizontal direction. The dust particles filled only a single horizontal layer, and significant out-of-plane displacements~\cite{Samarian:01} were not present, so that particles within the dust layer can be described as having mainly two-dimensional (2D) motion. As a single dust particle moved at a velocity ${\bf V}_d$, it experienced a frictional drag force $- \nu_{\rm gas} m_d {\bf V}_d$ due to collisions with the gas atoms~\cite{Liu:03}, where $\nu_{\rm gas} = 2.7~{\rm s^{-1}}$. The dust particles had an interparticle distance $b = 0.50~{\rm mm}$, corresponding to a Wigner-Seitz radius~\cite{Kalman:04} $a = 0.26~{\rm mm}$. At this interparticle distance, the longitudinal sound speed~\cite{Feng:10} in the crystalline lattice was measured to be $\approx 25~{\rm mm/s}$.

We generated a stable subsonic laminar flow pattern analogous to a Couette flow. In a traditional Couette flow, which fills a 3D volume, a pair of planar plates manipulate a fluid between them by moving at a steady speed in the $x$ direction, so that the flow velocity $v$ has only a single vector component, $v_x(y)$, which varies only with the coordinate $y$ perpendicular to the flow~\cite{Liepmann:57}. The shear $dv_x/dy$ is responsible for the viscous heating in Eq.~(\ref{energy}).

In our experiment, instead of planar plates we used laser manipulation~\cite{Homann:98, Juan:01, Liu:03, Nosenko:04, Wolter:05, Vaulina:08, Fink:11} of the dust particles to cause them to flow. Laser beams can drive significant flows, with strong gradients, due to the extreme softness of the dusty plasma. Using a rapidly-rastered scanning mirror, we dispersed a 532-nm laser beam (with a power of $2.28~{\rm W}$ as measured inside the chamber) into a sheet of width approximately $4~{\rm mm}$, which was incident on the dust layer at a downward angle of $6^\circ$. This laser sheet drove a flow in the $+x$ direction. A second laser sheet, separated by about $5~{\rm mm}$ from the first, was directed oppositely to drive a flow in the $-x$ direction, as shown in Fig.~1. Using laser sheets of $\approx 4~{\rm mm}$ width (instead of narrowly-focused laser beams) allowed us to generate wider flow velocity profiles. The video can be seen in the Supplemental Material~\cite{video}. Our shear region allowed us to measure the velocity profile $v_x(y)$ and the kinetic temperature profile $T_{\rm kin}(y)$ with sufficient spatial resolution to characterize the three dimensionless parameters: ${\rm Br}$, ${\rm Pr}$, and ${\rm Ec}$.

To determine the flow velocity ${\bf v}$ and kinetic temperature $T_{\rm kin}$ of the flowing dust particles as functions of $y$, we used video imaging and particle tracking~\cite{Feng:07, Feng:11_2}. The dust particles were imaged from above at 55 frames per second, and in each video frame we analyzed the particles within a region of interest measuring $23.5~{\rm mm} \times 23.5~{\rm mm}$ and containing $\approx 2500$ dust particles. We measured~\cite{Feng:07} the velocity ${\bf V}_d$ of individual dust particles. We then converted these particle data to continuum data by a spatial average of the particle data within rectangular bins of width $\Delta y = 0.26~{\rm mm}$ extending across the full region of interest~\cite{Feng:10}. In doing this, we assumed that $x$ is an ignorable coordinate in the flow, which we have validated~\cite{Feng:12}, and that our results for ${\bf v}$ and $T_{\rm kin}$ are functions only of $y$~\cite{continuum}. We used the particle velocities ${\bf V}_d$ to calculate the flow velocity ${\bf v} = \overline {{\bf V}_d}$ and the kinetic temperature $T_{\rm kin}(y) = m_d \overline{|{\bf V}_d - {\bf v}|^2 }/2k_B$, where the overline indicates a spatial average. We note that the kinetic temperature, which is calculated using the mean-square velocity after subtracting the local flow velocity, is not necessarily the same as a thermodynamic temperature $T$.

\begin{figure}[htb]
\centering
\includegraphics{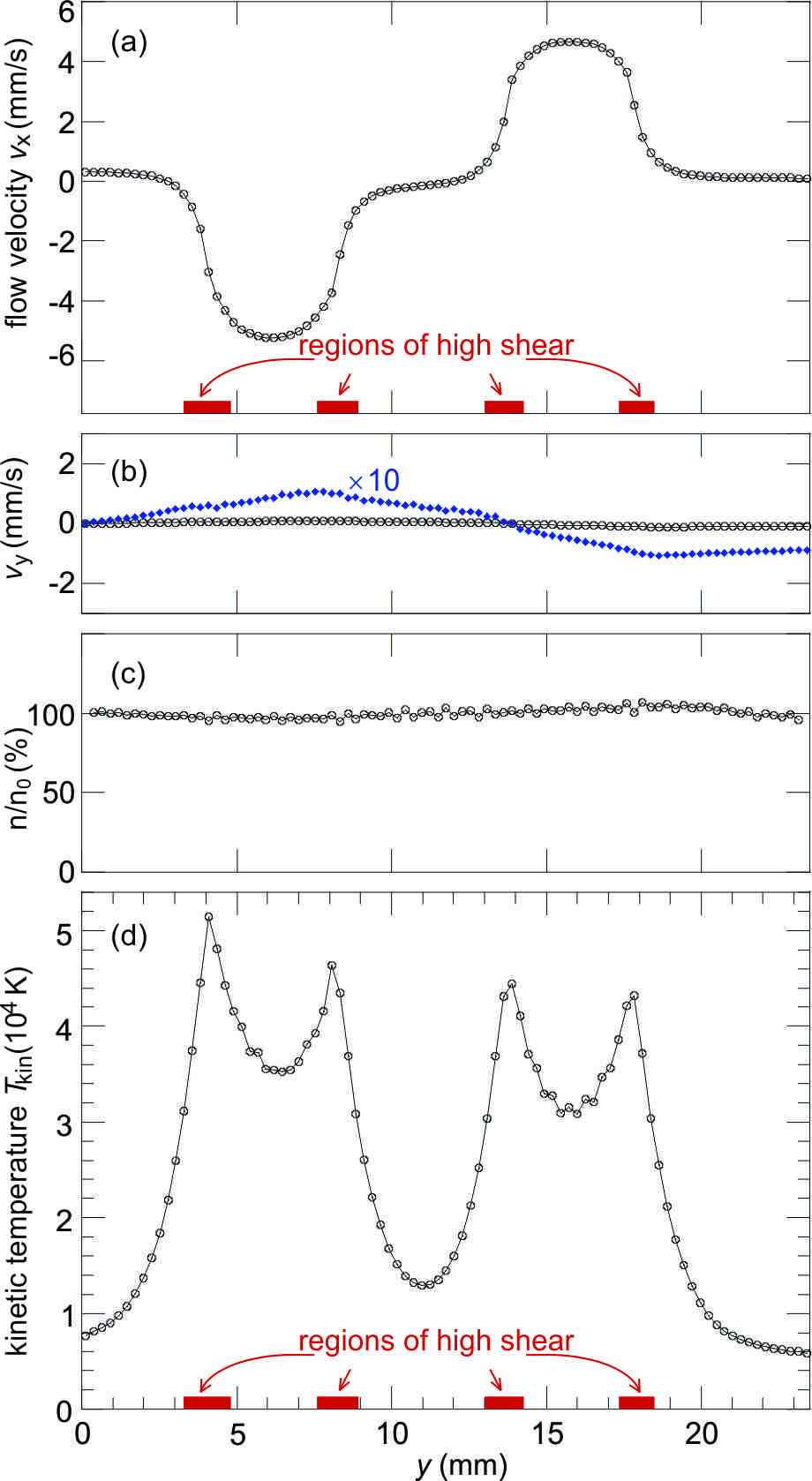}
\caption{\label{etasketch} (Color online). Spatially-resolved profiles of the continuum quantities. The two broad peaks in the profile of $v_x$ in (a) are due to the laser manipulation. We are interested mainly in the regions of high shear, $dv_x/dy$. The symmetry of the experiment minimized any flows in the $y$ direction, (b), and the number density was nearly uniform, (c). One of our chief results is the appearance of profound temperature peaks, (d). This kinetic temperature, shown in Kelvin, was determined directly from the motion of individual dust particles. The profiles of $v_x$ and $T_{\rm kin}$ allow us to determine Br and Ec, and verify that viscous heating has a strong effect. Marked at the bottom are the regions of high shear, where $|dv_x/dy| \ge 1.3~{\rm s^{-1}}$.}
\end{figure}

In our results for the velocity profile, Fig. 2(a), we see broad peaks, which are laser-driven flows. On each side of the broad peaks there is a region of high shear, where the gradient $dv_x/dy$ reaches its maximum of $4.3~{\rm s^{-1}}$. Most of our attention in this Letter is devoted to these shear regions.

Our experiment achieved a close approximation to the symmetry of a Couette flow, which would ideally have $v_y / v_x = 0$ everywhere and a uniform number density. In our experiment, $|v_y / v_x | \approx 10^{-2}$  as can be seen by comparing panels (a) and (b) of Fig.~2, and the number density in Fig.~2(c) is uniform within our measurement uncertainties. We note that unlike a traditional Couette flow for a single-phase liquid, ours has a nonlinear velocity profile $v_x(y)$; this is due to frictional drag applied to the dust particles by the gas~\cite{Nosenko:04}.

The first chief result of this Letter is the appearance of profound peaks in the kinetic temperature profile, Fig.~2(d). These peaks coincide with the locations of maximum shear, $|dv_x/dy|$. Unlike most cases where viscous heating in a shear flow leads to only a tiny change in temperature, here we see a very large effect, with a six-fold change in absolute temperature over only a few mm. We will discuss the significance of these temperature peaks below.

In the corresponding video of the particle motion~\cite{video}, one can see the effect of viscous heating in the motion of individual particles. In the high shear region, there are frequent collisions of particles at slightly different values of $y$, so that they are flowing at different speeds, and these collisions cause a scattering of momentum and energy that increases the random fluctuations of particle velocities. These particle velocity fluctuations correspond directly to a high kinetic temperature. This is the viscous heating effect, as observed at the particle level.

In order to achieve large values of Br and Ec, it is desirable to have a velocity shear that is large in dimensionless units. To make it dimensionless, we divide $dv_x/dy$ by $v_{th} / a$, where $v_{th} = \sqrt{2 k_B T / m_d}$ is the thermal velocity and $a$ is the the Wigner-Seitz radius. The dimensionless shear $(dv_x /dy)/(v_{th} / a)$ in our experiment is $\approx 1$. This is a large value; for comparison, the dimensionless shear is about $10^{-4}$ for a laminar boundary layer of air flowing at $100~{\rm m/s}$ over a cm-size flat plate~\cite{Blasius}.

To determine the three dimensionless numbers (Br, Pr, and Ec), we require not only the spatial profiles of flow velocity and temperature in Fig.~2, but also values for transport coefficients $\eta$ and $\kappa$. We obtained these coefficients by minimizing residuals of both the energy equation, Eq.~(\ref{energy}), and the momentum equation, while treating $\eta$ and $\kappa$ as free parameters. The residual is an equation's right-hand-side, which is finite (instead of zero, as in the ideal case) when using an input of experimental data. Details of this residual-minimization method are presented elsewhere~\cite{Feng:12}, where we obtained our results for the kinematic viscosity and thermal diffusivity, $\nu  = \eta / \rho  = 0.69~{\rm mm^2/s}$ and $\chi = \kappa / (c \rho) = 8~{\rm mm^2/s}$, respectively. These values, which were obtained simultaneously from the same experiment, are consistent with values reported for $\nu$ in a previous dusty plasma experiment~\cite{Nosenko:04} and $\chi$ in another~\cite{Nosenko:08}.

As the second chief result of this Letter, we now determine three dimensionless numbers for the flow in our dusty plasma.

The Prandtl number ${\rm Pr}$ is an intrinsic property of a substance. We obtain its value by evaluating Eq.~(\ref{Prandtl}) using the values for $\nu$ and $\chi$ given above. We find ${\rm Pr} = \nu/\chi = (0.69~{\rm mm^2/s}) / (8~{\rm mm^2/s}) = 0.09$ for our dusty plasma. For comparison, typical values for air~\cite{Handbook} and glycerin~\cite{White:00} are ${\rm Pr} \approx 0.7$ and 11000, respectively.

The Eckert number ${\rm Ec}$ characterizes dissipation in a flow. Unlike ${\rm Pr}$, it is not an intrinsic property of materials, but depends on the flow profiles. We obtain its value using Eq.~(\ref{Eckert}) with data from Figs.~2(a) and 2(d). Examining Fig.~2(a), we find $\Delta v = 2.5~{\rm mm/s}$, where we specified $\Delta v$ as the difference of the flow velocity between the positions $y = 8.3~{\rm mm}$ and $y = 11.0~{\rm mm}$, corresponding to maximum shear $|dv_x/dy|$ and minimum speed $|v_x|$, respectively. Using Fig.~2(d), we obtain the kinetic temperature difference $\Delta T_{\rm kin}$, specified for the same two positions. From Fig.~2(d) we find $\Delta T_{\rm kin} = 4.6 \times 10^4 \, {\rm K} - 1.3 \times 10^4 \, {\rm K} = 3.3 \times 10^4 \, {\rm K}$. We then  calculate $c\,\Delta T_{\rm kin} = \Delta(v_{th}^2/2) = k_B\,\Delta T_{\rm kin} / m_d = 1.1~{\rm mm^2/s^2}$. Inserting these values in Eq.~(\ref{Eckert}), we obtain ${\rm Ec} = 5.7$. This large value of Ec indicates that in our dusty plasma flow the transfer of directed kinetic energy to thermal energy is significant.

The Brinkman number, which characterizes the competition between viscous heating and thermal conduction for a given flow, is obtained simply by multiplying the Prandtl and Eckert numbers, ${\rm Br} = {\rm Pr}\,{\rm Ec}$. Using the values given above, we find ${\rm Br} = 0.5$ for the flow in our dusty plasma. This large value indicates that the first term in Eq.~(\ref{energy}) is generally half as large as the second term, i.e., viscous heating is half as large as thermal conduction. The appearance of the temperature peaks in the region of sheared flow in Fig.~2(d) is consistent with the large value of Br. In our search of the literature mentioning the Brinkman number, we found that temperature peaks have certainly been predicted theoretically, for example in Couette flows~\cite{Mills:95}, but we found no reports of an experiment with a spatially-resolved {\it in-situ} temperature measurement like ours. The microscopic mechanism for viscous heating is scattering of momentum by collisions of particles flowing at different speeds in neighboring portions of a sheared flow, which introduces random kinetic energy. In many substances, this local increase in random kinetic energy would be rapidly transported away by thermal conduction, so that no significant local heating of the substance can be observed. In our dusty plasma, however, temperature peaks in the region of maximum shear are easily observed, indicating that thermal conduction is not sufficient to flatten the temperature profiles.

To demonstrate the significance of viscous heating in our dusty plasma, we can compare it with other substances using the Brinkman number. For water flowing in a channel with a width of about $1~{\rm mm}$ (which is comparable to the scale lengths in our experiment), Tso and Mahulikar~\cite{Tso:00} found ${\rm Br}$ in the range  $1 \times 10^{-8}$ to $17 \times 10^{-8}$. Such low values indicate that viscous heating causes very little temperature change, due to the strong competing effect of thermal conduction. For glycerin in a Taylor-Couette flow, White and Muller found ${\rm Br} = 0.0359$~\cite{White:00}. For a polymer flowing as it is squeezed in an extruder used for manufacturing thermoplastics, where viscous heating is important because it melts the substance, Housz and Meijer~\cite{Housz:81} found ${\rm Br} = 0.47$. In our literature search, the only experiment that we found with a Br larger than ours was performed using an exotic liquid that requires a different definition of Br~\cite{Sukanek:74}. We have also not found any previous experimental study of viscous heating that includes a measurement of the temperature profile, revealing the spatial localization of viscous heating, as we have done in Fig.~2(d).

In summary, using {\it in-situ} measurements of the kinetic temperature in a highly sheared flow of a dusty plasma, we have demonstrated that viscous heating can result in profound temperature peaks. These temperature peaks coincide with the locations of maximum shear. We determined Br = 0.5, Pr = 0.09, and Ec = 5.7. Our large value of Br indicates viscous heating to an extent that is seldom measured experimentally in other substances. Further details of the experiment, including spatially resolved determinations of the terms in the continuum equations for momentum and energy, will be presented in another paper~\cite{Feng:12}.

This work was supported by NSF and NASA.


\begin{thebibliography}{16}
\bibitem{continuum}
We also assume that it is valid to use a continuum model when gradients are concentrated in a region as small as a few particle spacings. The validity of this assumption was demonstrated for 2D dusty plasmas experiments in~\cite{Nosenko:04, Nosenko:08}.
\end{thebibliography}
\end{document}